\documentclass{article}
\usepackage{amssymb}

\usepackage{graphicx}
\usepackage{epsfig}

\begin{document}

\title{Capacity formulas in MWPC: some critical reflexions.}

\author{P. Van Esch}

\maketitle

\begin{abstract}
An approximate analytical expression for "capacitance" of MWPC
configurations circulates in the literature since decades and is
copied over and over again. In this paper we will try to show that
this formula corresponds to a physical quantity that is different
from what it is usually thought to stand for.
\end{abstract}

\section{Introduction}

Simple as it may seem, the concept of capacitance is a many-faced
item when working in a setup with many electrodes such as it is
the case in a MWPC, and it is easy to get lost.  We will first
analyze exactly what are the different accepted definitions of
capacitance and how they are related to physical quantities that
affect the functioning of the detector.  Next we compare that to
the established formula we will analyze here.  It is the
expression for "the capacitance per unit length of each wire" in
the case of a parallel wire grid, symmetrically placed between two
conducting planes:
\begin{equation}
\label{eq:sauliformula} C = \frac{2 \pi \epsilon_0}{\frac{\pi
l}{s}- \ln \frac{2 \pi r_s}{s}}
\end{equation}
In this expression, $s$ is the spacing of the wires in the grid,
$l$ is the distance between the wire plane and each of the
conducting planes and $r_s$ is the radius of the wires.  The
formula is based upon an approximation which is excellent when $s$
is of the order of or smaller than $l$.  This equation can be
found in \cite{saulimwpc} and in about all courses on gas
detectors.  It finds its origin in a formula in an old paper by
Erskine \cite{erskine}, where Erskine calculated the accumulated
charge on a wire plane between two planar conductors. The problem
is that stated this way, the formula seems to imply that the above
quantity called $C$ is, for instance, the capacitance as seen by
the input of an amplifier that is connected to a single wire. This
is the capacitance that will then determine the noise current
induced by the equivalent voltage noise (series noise) of the
amplifier. We will show in this paper that that is not true: $C$
is not that quantity.  But one can already see that there is
something disturbing about the given formula: namely the fact that
the capacitance per wire {\it increases} when, all other
dimensions equal, the wire spacing increases. Is equation
\ref{eq:sauliformula} wrong then ? The answer is no.  The formula
does describe a quantity that can be called a capacitance, but it
is not the usual definition --- and it is not what the amplifier
will see at its entrance.

\section{Definitions of capacitance.}

In order to understand the different possible definitions of
capacitance and the confusion it can lead to, a short, elementary
review is due. One cannot do better than to return to Jackson
\cite{jackson} in order to have a clear definition of what is
"capacitance". There, on p. 43, it is clearly stated: "For a
system of $n$ conductors, each with potential $V_i$ and total
charge $Q_i$ in otherwise empty space, the electrostatic potential
energy can be expressed in terms of the potentials alone and
certain geometrical quantities called coefficients of capacity.",
and further:
\begin{equation}
\label{eq:capacitancematrix} Q_i = \sum_{j=1}^n C_{ij}V_j
\end{equation}
We could even add to this that "empty space" can be a confined
volume with an enclosing conducting wall at ground potential.
Next, Jackson writes: "The coefficients $C_{ii}$ are called
\textbf{capacitances} while the $C_{ij}, i\neq j$ are called the
\textbf{coefficients of induction}." and "The capacitance of a
conductor is therefore the total charge on the conductor when it
is maintained at unit potential, all other conductors being held
at zero potential.". Note that capacitance is normally a positive
quantity, while the coefficient of induction is normally a
negative quantity.  The coefficient of induction $C_{12}$ is the
"crosstalk" capacitance which induces charges on conductor 1 when
voltages (with respect to ground) appear on conductor 2, all other
conductors, including conductor 1, remaining at the same potential
(with respect to ground).

It is the capacitance, as defined by Jackson, that is "seen" by
the input of an amplifier (and hence enters into the noise
calculations), when all conductors are connected to
(low-impedance) charge amplifiers.

Jackson also defines: "\textbf{the capacitance of two conductors}
carrying equal and opposite charges in the presence of other
grounded conductors is defined as the ratio of the charge on one
conductor to the potential difference between them".  This can
then easily be worked out to result in the following expression:
\begin{equation}
C_{1-2} = \frac{C_{11} C_{22} -
C_{12}C_{21}}{C_{11}+C_{22}+C_{12}+C_{21}}
\end{equation}
which reduces, in the symmetrical case ($C_{11} = C_{22}$,
$C_{12}=C_{21}$), to:
\begin{equation}
C_{1-2} = \frac{C_{11}-C_{12}}{2}
\end{equation}
This is the capacitance that is measured by a floating capacitance
meter between conductors 1 and 2 (when all other conductors are
put to ground potential).  Note that numerically, $C_{1-2}$ is
bigger than $|C_{12}|$, because it includes also the indirect
capacitive coupling in series: node 1 - ground - node 2. For
instance, if there is no direct coupling ($C_{12}=0$) we find,
indeed, that $C_{1-2} = C_{11}/2$.

\begin{figure}
\centering
  \includegraphics[width=8cm]{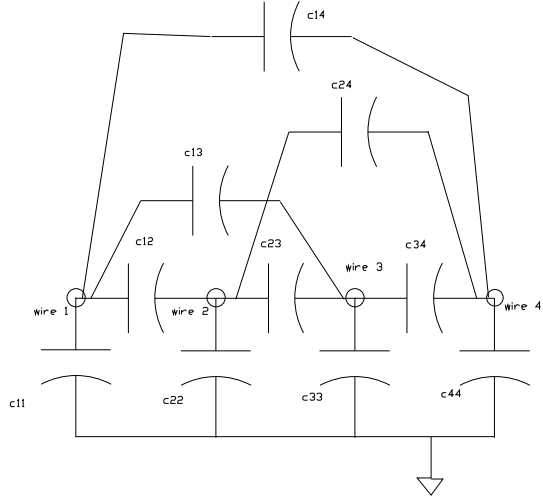}\\
  \caption{Equivalent capacitor network.}\label{fig:equinet}
\end{figure}

What is the relationship between these quantities and a network of
"equivalent capacitors" linking all conductors (nodes) amongst
them and to ground, as shown in figure \ref{fig:equinet} ?  Let us
note by $c_{ij}, i \neq j$ the equivalent capacitor linking nodes
$i$ and $j$, and by $c_{ii}$ the equivalent capacitor linking node
$i$ to ground. We now have a passive linear network to which we
can apply the well-known method of node potentials
\cite{schaumelectricity} to write (with $p$ the Laplace variable):
\begin{equation}
\left(%
\begin{array}{ccc}
  p\sum_{k=1}^n c_{1k} & -p c_{12} & ... \\
  -p c_{21} & \sum_{k=1}^n c_{2k} & ... \\
 ... & ... & ... \\
\end{array}%
\right) \times
\left(%
\begin{array}{c}
  v_1 \\
  v_2 \\
  ... \\
\end{array}%
\right)
=
\left(%
\begin{array}{c}
  i_1 \\
  i_2 \\
  ... \\
\end{array}%
\right)
\end{equation}

Bringing $p$ to the other side, we obtain, on the right hand side,
elements of the form $i_1/p$ which, in the time domain, come down
to integrating the current over time, so $i_1/p$ can be replaced
by the charge $Q_1$ etc... and we recognize the equivalences
between the capacitance matrix of Jackson and the elements of the
node voltage conductance matrix above:  $c_{ij} = - C_{ij}, i\neq
j$ and $c_{ii} = \sum_{k=1}^n C_{ik} = C_{ii} - \sum_{i \neq
j}|C_{ij}|$.

\section{The meaning of Erskin's formula}

Equation \ref{eq:sauliformula} is based upon an expression Erskin
derives in \cite{erskine}, when he calculates an approximate
expression for the charge $Q$ on each wire when $\it all$ wires
are brought to a potential $V_0$. Formula \ref{eq:sauliformula} is
then nothing else but the ratio of $C = Q/V_0$. Let us consider an
arbitrary wire number 1 ; using equation
\ref{eq:capacitancematrix}, we can then easily derive that $C =
\sum_{k=1}^n C_{1k}$ and this is equal to $c_{11}$: it is the
capacitor element from the wire to ground in the equivalent
network. But note that this is NOT the capacitance of the wire
with respect to ground which is $C_{11}$.

Unfortunately, the only way to measure, in a direct way, $c_{11}$,
is by connecting all other wires to the output of a 1:1 buffer
amplifier which has a high impedance and whose input is connected
to wire number 1, using all other wires in an active shielding
configuration. If we then measure, with a capacitance meter, the
capacitance of wire 1 w.r.t. ground, we will find $c_{11}$.
Indeed, the only capacitor on which the charge can flow is on
$c_{11}$: all $c_{1k}$ capacitors are, through the servo mechanism
of the buffer amplifier, kept on the same potential on both sides
and do not take in any charge. This also explains the
counterintuitive behavior of equation \ref{eq:sauliformula}, that
when the wires get closer, $C$ diminishes: indeed, there is more
and more active shielding of the ground plane by the nearby wires,
and less and less direct coupling to the ground plane, so the
closer the active shielding wires come, the less capacity is
measured.

However, it is now also clear that this "capacitance" $C$ is
almost never the physical quantity we need in an actual
application (such as the load to the entrance of an amplifier).
Only in the limit of large $s$, when $c_{12}$ goes to 0, would
$c_{11}$ become equal to $C_{11}$, but there the approximation
used is not valid anymore !

\begin{figure}
\centering
  \includegraphics[width=8cm]{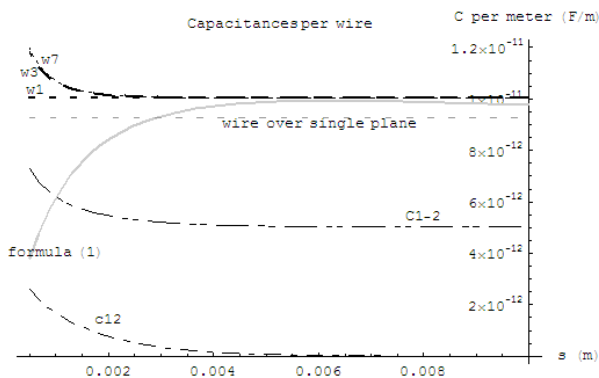}\\
  \caption{Capacitance of a single wire, calculated in different ways:
  w1, w3 and w7 stand respectively for the $C_{11}$ calculation with 1,
  3 or 7 wires.  They are compared the formula \ref{eq:sauliformula}, and
  to the formula for the capacitance of a wire over a single plane.
  $c_{12}$, the induction coefficient, and $C_{1-2}$, the
  capacitance between two adjacent wires, are also displayed.}\label{fig:eigenvalues}
\end{figure}

If we connect the other wires to a high-impedance (voltage)
amplifier (essentially leaving them floating) this comes down to
having no possibility of having a flow of net charge on these
wires when wire 1 is brought from 0 to 1 V.  In this case, it is
as if these wires are absent, and the capacitance measured on wire
1 will be the capacitance of a single wire (limit $s \rightarrow
\infty$), which is represented by the line w1 in figure
\ref{fig:eigenvalues}.  As such, we should have a capacitance
which is independent of $s$; again not the value given by $C$. In
fact, it is impossible without using active components to make the
value of the capacitance as seen by an amplifier descend below the
value of w1. Any passive load on the other wires will $increase$
the capacitance seen by the amplifier, through $c_{12}$, and not
decrease it, as does formula \ref{eq:sauliformula}.

We can compare the result of an exact calculation of the
capacitance $C_{11}$ of wires of 1 m length (in a semi-analytic
way, in a very similar way as done by Erskine \cite{erskine}) of
the middle wire of a set of 1, 3 or 7 wires (curves w1, w3 and w7)
with equation \ref{eq:sauliformula}. We also include a calculation
of the capacitance of a single wire over a ground plane. We take
the case of wires with a diameter of 20 $\mu m$, a distance
between the wire plane and each of the ground planes of 2mm and we
plot the quantities as a function of the wire spacing.  We also
show the values of $C_{1-2}$ and of $c_{12}$ for comparison.

This is shown in figure \ref{fig:eigenvalues}
Clearly, although in a certain range, by coincidence, the
numerical values of both calculations are of the same magnitude
(and of the order of $\epsilon_0$), both curves have nothing to do
with one another.  The value needed in most applications is not
the one given by formula \ref{eq:sauliformula}.

\section{Discussion}

In this paper we reviewed the different aspects of the concept of
"capacitance" and used this to confront it to the calculation of a
"standard formula" for a "capacitance" $C$, equation
\ref{eq:sauliformula}, well-known in the world of gas detectors.
From this comparison, it turns out that this quantity has a
meaning, namely a capacitor value in the equivalent circuit
describing the capacitive interactions between the wires and the
ground plane $c_{11}$, but that this is not the quantity it is
usually claimed it is supposed to be (namely the capacitance of a
single wire $C_{11}$). The confusion between both quantities
(which are shown to have numerically different behavior) can lead
to wrong applications and wrong conclusions, for instance,
concerning the noise behavior of amplifiers connected to the wires
of a MWPC.

\end{document}